\documentclass[acmsmall]{acmart}

\usepackage[colorinlistoftodos]{todonotes} 
\usepackage{setspace}
\usepackage{tabularx}
\usepackage{enumitem}
\usepackage{multirow}
\def\BibTeX{{\rm B\kern-.05em{\sc i\kern-.025em b}\kern-.08emT\kern-.1667em\lower.7ex\hbox{E}\kern-.125emX}}

\setcopyright{acmlicensed}
\acmJournal{PACMHCI}
\acmYear{2019} \acmVolume{3} \acmNumber{CSCW} \acmArticle{119} \acmMonth{11} \acmPrice{15.00}\acmDOI{10.1145/3359221}
\copyrightyear{2019}
\acmConference[CSCW '19]{ACM Conference on Computer Supported Cooperative Work and Social Computing}{November 09--13, 2019}{Austin, TX}
\acmBooktitle{CSCW '19: ACM Conference on Computer Supported Cooperative Work and Social Computing, November 09--13, Austin, TX}

\acmSubmissionID{cscw119}

\begin{document}

%
\title{This Thing Called Fairness: Disciplinary Confusion Realizing a Value in Technology}
%

\author{Deirdre K. Mulligan}
\affiliation{
\institution{School of Information, University of California, Berkeley}
\country{USA}
}
\email{dmulligan@berkeley.edu}

\author{Joshua A. Kroll}
\affiliation{
\institution{School of Information, University of California, Berkeley}
\country{USA}
}
\email{kroll@berkeley.edu}

\author{Nitin Kohli}
\affiliation{
\institution{School of Information, University of California, Berkeley}
\country{USA}
}
\email{nitin.kohli@ischool.berkeley.edu}

\author{Richmond Y. Wong}
\affiliation{
\institution{School of Information, University of California, Berkeley}
\country{USA}
}
\email{richmond@ischool.berkeley.edu}

%
\renewcommand{\shortauthors}{Mulligan, Kroll, Kohli, and Wong}

%
\begin{abstract}
The explosion in the use of software in important sociotechnical systems has renewed focus on the study of the way technical constructs reflect policies, norms, and human values. This effort requires the engagement of scholars and practitioners from many disciplines. And yet, these disciplines often conceptualize the operative values very differently while referring to them using the same vocabulary. The resulting conflation of ideas confuses discussions about values in technology at disciplinary boundaries. In the service of improving this situation, this paper examines the value of shared vocabularies, analytics, and other tools that facilitate conversations about values in light of these disciplinary specific conceptualizations, the role such tools play in furthering research and practice, outlines different conceptions of ``fairness''deployed in discussions about computer systems, and provides an analytic tool for interdisciplinary discussions and collaborations around the concept of fairness. We use a case study of risk assessments in criminal justice applications to both motivate our effort--describing how conflation of different concepts under the banner of ``fairness'' led to unproductive confusion--and illustrate the value of the fairness analytic by demonstrating how the rigorous analysis it enables can assist in identifying key areas of theoretical, political, and practical misunderstanding or disagreement, and where desired support alignment or collaboration in the absence of consensus.
\end{abstract}

%
%

%
\keywords{analytic; fairness; values; infrastructure; interdisciplinary}

%

%
\maketitle
\section{Introduction}
Governments, companies, and the technologists they employ are increasingly being called to task for the social consequences of the technologies they build and use.
Architectures, interfaces, configurations, defaults, and algorithms must yield their biases and impacts on values such as privacy, equality, security, freedom of expression, and access to information.
This increased focus on the values reflected in the technical artifacts themselves and the larger organizational structures they reconfigure rests on the insight that technology is not merely a tool of implementation, but constitutive of an organization's values commitments in practice. 
In particular, a growing body of work focuses on how to produce technical artifacts and organizational structures that promote values such as fairness, accountability, transparency, and explainability in software systems.

This keen focus on the values implications of technological choices, both independently and relationally, is a break with past practice where biases and values were largely viewed as exogenous to technical design, and addressed through the adoption of legal rules and institutional policies and processes.
Academic traditions of interrogating the values of technical and sociotechnical systems have deep roots from multiple disciplines, including science \& technology studies, law, philosophy, HCI, and computer science \& engineering  (e.g., \cite{latour1989moral, latour1992missing,  clark2002tussle, lessig1999code, reidenberg1997lex, rfc6973, friedman1996bias, Winner1980}).
In addition to the critical analysis of understanding values in technical artifacts, a body of research has also articulated a constructive agenda to build artifacts that promote or embed particular social values \cite{friedman1996value,nissenbaum2005values,flanagan2005values}.

This move from critical interrogation of values in technical artifacts and systems to a constructive agenda is arduous and requires \emph{sustained work across disciplines}~\citep{flanagan2008embodying}.
Advancing a constructive approach to building values into technical systems requires shifts in some disciplines' orientation toward values. 
It requires an openness in the technical community to view values as endogenous or at least co-constituted by system design.
It requires individuals with the disposition, skill, resources, and intellectual breadth to bridge different disciplines.
Constructive work to embed values also requires new tools.
Researchers and practitioners from disparate disciplines need conceptual and methodological tools that allow them to flesh out the problem, develop a research agenda, and work with relevant values---the ``values at play''~\citep{flanagan2005values}.
These tools are an integral component of the infrastructure necessary to support cross-disciplinary work.\footnote{
We use the term cross-disciplinary throughout as Paul Jeffrey does ``to mean all forms of collaboration between researchers with different educational backgrounds''~\citep{jeffrey2003smoothing}.
}

In this paper we use the term ``infrastructure'' in two ways. First, this paper helps to build and support the research infrastructure necessary to support the work of the interdisciplinary group of researchers and practitioners in the emerging loosely coordinated community of technologists, social scientists, lawyers, and policymakers studying fairness, accountability, transparency, explainability, and other values in software systems. We provide a definitional overview documenting how ``fairness'' in this emerging field is used across the disciplines represented.
We also provide a Fairness Analytic, a set of dimensions to discuss or think through what is meant by ``fairness'' across multiple conceptions or disciplines. 
These are first steps toward the development of a shared mapping of vocabularies, and a tool to aid researchers in avoiding misunderstanding, or to engage in more constructive debates of difference.

Second we use ``infrastructure'' as a conceptual lens, going beyond referring to social and technical resources and artifacts. The analytical lens of infrastructures asks us to pay attention to the \textit{work and practices} done in the ``background'' that maintain and support sociotechnical systems  \cite{bowker2009toward, star1994steps, Rosner2014Designing}. 
This lens also recognizes that the practices, resources, and artifacts that make up an infrastructure are themselves values-laden and political. 

Vocabularies scaffold research; yet like other infrastructure, they often go unnoticed despite their essential contribution to interdisciplinary work~\citep{star1994steps}.
Below we describe the importance of cross-disciplinary infrastructure; describe five ways shared vocabularies support productive cross-disciplinary work; offer a definitional overview of concepts of fairness, examining how the term is defined and conceptualized in different disciplines; provide a Fairness Analytic to provide a set of dimensions to discuss concepts fairness across definitions and disciplines; apply these tools to a case study; and reflect on this work as an incremental step toward building a shared vocabulary that fosters the collaborative work of this emerging community.
While the community focused on the terms we consider here is at the early stages of formation, now is the time to attend to the infrastructure necessary to support its development.

\section{Background}
Particularly with the rise of machine learning as a core technical tool for building computer systems and the concomitant sense that these systems were not adequately reflecting the values and goals of their designers and creators, the question of how to build values into software systems has gained significant traction in recent years.
One focal point for the community has been the rise of the FAT scholarly meetings (FAT/ML, or the Workshop on Fairness, Accountability, and Transparency in Machine Learning, held annually since 2014 at a major machine learning conference and FAT$^*$, the (now, ACM) Conference on Fairness, Accountability, and Transparency, which aims to build community outside of research with a machine learning focus.\footnote{See https://fatml.org and https://fatconference.org. This area has seen explosive growth in research activity: major academic institutions have convened cross-disciplinary scholars around this topic in events and internal study groups. Consider New York University's Algorithms and Accountability (http://www.law.nyu.edu/centers/ili/algorithmsconference) and Algorithms and Explanations (http://www.law.nyu.edu/centers/ili/events/algorithms-and-explanations) conferences, the Yale Information Society Project's ``Unlocking the Black Box'' conference (https://law.yale.edu/isp/events/unlocking-black-box), and UC Berkeley's Algorithmic Opacity and Fairness Working Group (https://afog.berkeley.edu/), to present only a few examples of strong interdisciplinary focus.}

In a similar vein, prior research within Computer Supported Cooperative Work (CSCW) and related fields, such as Human Computer Interaction (HCI) and Science \& Technology Studies, has paid attention to the ways in which technical practices and computational artifacts of all kinds embed or promote a range of social values (e.g., \cite{Shilton2014, Winner1980, Knobel2011, Nissenbaum2001, Shilton2018}). 
Research programs developed to focus on values and technology, such as Value Sensitive Design and ``values in design'', include forms of both analysis to identify and critique values associated with systems, and methods for incorporating values into the processes of engineering and design \citep{friedman1996value,Friedman2008,Knobel2011,Shilton2018}.  
Some of these approaches to values have been critiqued as viewing values as overly stable and universal, without always accounting for how values arise locally in situated environments \cite{LeDantec2009, JafariNaimi2015}.

Recent research has moved away from considering values as universal and stable.
Instead, values are seen as instantiated through specific situated practices \citep{JafariNaimi2015,Houston2016,LeDantec2009}.
The conceptions of values presented in our definitional overview reflect this tradition.
Rather than argue for a single ``proper'' conception of fairness,  we instead use the analytic to recognize and connect the diversity of ways these terms are defined and operationalized within different disciplinary and professional contexts, and different communities of practice including those of law, social sciences, humanities, and computer science.
In particular, examining discussion around and representation of values like fairness in law and policy discussions as well as in social science and technical discussions recognizes the mutual intersections and interactions, or ``knots'' \citep{Jackson2014Policy}, of these seemingly disparate fields.
This builds on work by researchers who have created a range of conceptual tools to support interdisciplinary collaboration and discussion of values, such as Shilton et al.'s values dimensions~\citep{Shilton2014}, which helps researchers think about values relationally---such as where or with whom do values reside, at what scale are values being analyzed or designed for, and so on.

A growing body of CSCW research intersects with the broader interdisciplinary community focusing on fairness, transparency, accountability, explainability, and other values in computational artifacts -- particularly focusing on how these values come to matter in the creation, use, deployment, and maintenance of algorithmic and machine learning systems.
Indicative of this interest, recent workshops in the CSCW community interrogate the diversity of ways in which humans (and their values) relate to algorithmic ecosystems \citep{Wagenknecht2016,Wolf2018}.
The scope of this work ranges from empirical analysis to design, including: critical analysis of harms such systems can cause (particularly when purportedly recognizing and categorizing gender) and the violence they can perpetuate \citep{Keyes2018, hoffmann2017data, buolamwini2018gender}; ethnographic investigation of how data science and algorithmic work is taught, learned, and enacted in practice \citep{Passi2017,Passi2018}; user perceptions of ``fairness'' \citep{lee2017algorithmic}; and methods to design algorithms while being cognizant of stakeholder values \citep{Zhu2018}.

This paper helps strengthen this interdisciplinary community by illustrating the diversity of ways in which the concepts of values---specifically, fairness---have been promoted and maintained in different communities. 
We frame our analysis, the work involved in producing it, and potential diversity of ways it can be used through the analytical lens of infrastructures.

\section{The Need for Infrastructure}
In this paper, we first turn to the analytical lens of infrastructures, before discussing how that lens helps us develop a new piece of research infrastructure---the fairness analytic. 

Star and Ruhleder's question of ``When is an infratructure?'', rather than ``What is an infrastructure?,'' asks researchers to pay attention to the work and practices done in the ``background'' that maintain and support sociotechnical systems,  \cite{star1994steps}. 
A seemingly stable system--whether that be a bridge or a ``fair'' algorithm--only appears stable because of the the work and practices that people are doing to maintain and support it, which are referred to as practices of \textit{infratructuring} \cite{bowker2009toward}. 
The practices, processes, and tools used in infrastructuring are themselves values laden and political (e.g., who gets to make the decision about where the bridge should be built; who decides on what conception of fairness to use?).  

Reflecting diverse interests, stakeholders across the globe such as regulators, public interest advocates, and academics are demanding that computer systems be fair, transparent, accountable, respectful of human rights, and explainable.
A growing community of engineers and computer and information scientists, social scientists, humanities scholars are rising to the challenge.
The scientific interest in fairness, transparency, accountability, explainability, and other values in computer science has advanced the state of theory as well as application, engaging cutting edge research in cryptography~\citep{kroll2015accountable}, formal languages~\citep{albarghouthi2016fairness}, and systems~\citep{herschel2017survey,perez2018systematic}.
Research by social scientists and humanities scholars has similarly advanced the state of theory and application~\citep{barabas2017interventions, kohler2018eddie, binns2018s, wachter2019right, selbst2019fairness}.
But like other efforts to research, and address values in an integrative fashion, members of this community at times struggle to understand each other and it is unclear how their research outputs can yield systems that make progress toward the values-sensitive computer systems they imagine.

The conceptual lens of infrastructures helps provide several insights into fairness. 
First, terms like ``fairness'' that might at first glance seem stable, are only stable because people are doing work to create and maintain a particular conceptualization of fairness (such as by publishing research papers that define fairness in particular ways, or building systems that operationalize fairness in particular ways). 
Rather than view these practices of conceptualizing fairness as being in the ``background'' of designing and building systems, we can bring the practices and processes of conceptualization to the forefront and ask  \textit{how can we support the infrastructuring work of conceptualizing fairness}? 

The practices of processes involved in conceptualizing fairness is in part difficult and complex because it requires sustained work across disciplines. 
Sustaining the cross-disciplinary collaborations necessary requires more than scientific interest and advances, it requires attention to the people, processes, and tools that make ``fairness'' into something that multiple communities can grapple and work with.  
Our interest here is in highlighting the importance and politics of a particular component of  infrastructure---a shared analytic tool for fairness that helps a diverse research community identify and organize a research agenda and communicate their research both within the community and to external stakeholders despite different uses of the term fairness across disciplines and, at times, a lack of consensus within and across disciplines about what fairness requires in a given context.

Infrastructuring helps us see that the process of conceptualizing fairness does not just happen on its own--there is no single, universally ``natural'' conception of fairness. Rather, when the term ``fairness'' is used, its conceptual meaning exists because people have done social and technical work to create and maintain a particular conceptual meaning of fairness (through writing a definition, an equation, a line of code, and so forth). Our fairness analytic is an attempt to foreground the infrastructuring work implicit in current fairness activities. Through the analytic's structure, we provide a tool that exposes the politics of conceptualizing fairness and supports interdisciplinary engagements.

Moreover, the conceptual metaphor of infrastructure helps shed light on the ``messiness'' of fairness, which we keep in mind as we develop a piece of research infrastructure---a conceptual analytic to help discuss or think through what is meant by "fairness" across multiple conceptions or disciplines. 

\subsection{A Conceptual Analytic as Research Infrastructure}
\label{sec:analytic}
Cross-disciplinary work is difficult.
One noted barrier, among many, is the lack of a shared vocabulary.
As Paul Jeffries writes, ``Ideas generated within one discipline may make perfect sense within the discourse that they came from.
However, they will be evaluated in the context of acts and practices that do not function according to the discursive logic of the original discourse.
Hence, even though the representatives of different disciplines may be discussing a single, unambiguous topic, their vocabularies may be very different and mutually perplexing''~\citep{flanagan2008embodying}.
While divergent terms, and use of terms, across disciplines is nothing new, it presents a formidable barrier to progress where protection or support for the value in question is spread across the social, the legal, and the technical.\footnote{
Progress in the area of privacy by design has been hampered in part by the lack of conceptual and methodological tools that bridge disciplines.
A series of recent workshops held by the Computing Community Consortium trying to explore this space document these gaps: see~\citep{cccPBDworkshop1};~\citep{cccPBDworkshop2}; and~\citep{cccPBDworkshop3}.
}

Terms such as ``fairness'' and many others represent the linguistic conflation of disparate topics studied in different fields with different approaches, outlooks, methods and histories.
Even seemingly straightforward terms such as ``algorithm'', ``artificial intelligence'', and ``machine learning'' have murky boundaries and contested histories.
Existing terms come from situated and specific disciplinary contexts, histories and norms.
Advancements in the field require social scientists, humanists, and engineers to move beyond parallel play.
Tools that develop shared conversations and allow for meaningful contestation of ambiguous terms and the politics and assumptions that inform the various definitions at hand facilitate collaborative work by bridging disciplinary specific vocabulary.
Research at the interface of software systems and their human context (as well as practical policy making and specifically drafting and interpreting the law) necessarily engages concepts across disciplines.
However, because scholars and practitioners in different disciplines use the same words to mean different things, it can be challenging to advance understanding in a way that affects research or practice in different communities~\cite{holstein2019improving, kroll2019conceptions}.
Instead, faced with the question of how to describe concepts precisely, scholars and practitioners often double down on their existing disciplinary preconceptions, believing that resorting to their particular approach to rigor will surely convince those of different backgrounds.
Moving forward requires shifts in disposition toward other disciplines ways of seeing, as well as tools to explore the totality of approaches to seeing and describing the concepts at issue.

Analytic tools that tease apart complex concepts and bring disciplinary specific vocabularies that relate to them into conversation support productive engagement in at least four ways.
\begin{itemize}
\item First, the process of creating an analytic (which we present in Section~\ref{sec:taxonomy}) helps map the broad conceptual terrain afforded by different disciplines by offering a rich perspective on the problem and the potential solution space.
\item Second, conceptual analytics reduce needless arguments caused by loose use, under-specification, or competing definitions of terms. 
The community has already experienced this problem, as we will describe below with regard to rancorous debates about whether the COMPAS risk assessment was, or possibly could be, fair. 
These debates resulted in part from a lack of clarity about how fairness was being assessed.
\item Third, shared tools for analyzing ambiguous concepts allow researchers and others to understand how the insights and products of different fields relate, compare, and potentially compose into systems. This facilitates direct collaboration among researchers from different disciplines.
\item Finally, a shared conceptual analytic plays an important political role, in two distinct ways: 
It can foster \emph{political visibility}---that is, attention to the very existence and political nature of values questions being resolved by design choices~\cite{mulligan2018saving}---of the technical choices that enact often deeply contested values during research and system. And it also specifically protects against simplifications designed to make complex concepts tractable as engineering or math problems, or as being amenable to simple engineering solutions. 
\end{itemize}

\subsubsection{Collaboration}
Vocabularies both describe and define research problems.
The development of cross-disciplinary research agendas, and the research that follows, requires conceptual linkages across content areas which in turn depend upon tools that help a diverse community exchange knowledge in ways that support shared reasoning about a complex subject without requiring a full understanding of each discipline's work.

Tools to support a common understanding or discussion of terms are particularly important when working with concepts that are not only ambiguous within and across disciplines, but in some instances \emph{essentially contested}~\cite{gallie1955essentially}.
The terms this community is struggling collectively to advance---such as fairness---perform important, complicated, and at times quite distinct work in the realms of law, politics, markets, and system design~\cite{mulligan2018saving, kroll2017penn}.
What is fair in one may be far from fair in another; and even within one realm, ``fair'' can have multiple definitions, such as in law, which recognizes fairness in formally the application of procedures as well as substantively in the allocation of resources or punishments.

Opinions about what ``fair'' requires in a particular context are subject to heated debate, and understanding what is fair today may not capture what is considered fair tomorrow.
Nor is understanding what is considered fair by one stakeholder dispositive of what is considered fair by others.
For example, in relation to the concept of privacy, productive engagement with tangled, ambiguous and essentially contested concepts requires analytic tools for mapping the arguments, disputes and disagreements in which the concept arises~\cite{mulligan2016privacy}.
Thus we believe a conceptual analytic that teases out the dimensions of fairness (and eventually, other values) that are often essentially contested concepts is necessary to support collaborative work across disciplinary specific vocabularies around fairness in a manner that foregrounds the political, epistemological, contextual as well as practical aspects of choosing among competing definitions. The conceptual analytic, like a shared vocabulary, supports interdisciplinary collaborations, but importantly maintains the contested and dynamic aspects of values by tying language choice more tightly to assumptions about the world and knowledge production, and political commitments.

Given the multiple forms of diversity and distance represented in the research and practice communities around fairness, as well as its inherent contested-ness, a shared tool for reasoning may be quite useful. 
As with a previously proposed analytic for privacy~\citep{mulligan2016privacy}, the goal of such an effort is not to squelch or gloss over differences in our use of concepts, but to expose them to inquiry and facilitate the productive use of all the different concepts the respective disciplines have to offer in a rigorous and open way.
As ``the sophistication of $[\ldots]$ our scientific languages, is marked by $[\ldots]$ attunement to different possibilities of action to suit different needs and interests'' the effort to develop this analytic is pragmatic, an effort to build a tool to assist in ``coordinating our transactions with one another and our environment''~\citep{piso2016integration}.
Our aim is not an integration of the disciplinary perspectives that bear on the problem, but rather a tool that helps us see the value choices that rest in these perspectives so we can choose ones that align best with our goals.
This aligns with the tactic of ``infrastructural inversion,'' the surfacing of ways to organize the world (e.g., languages) for inquiry and potential new productive re-uses~\citep{bowker1994science, bowker2000sorting}.

\subsubsection{Political Visibility}
Next, a shared analytic for fairness supports collaborative work across disciplines and builds an infrastructure to maintain political visibility into technical choices~\citep{star1994steps}.
Infrastructures reflect communities' situated politics, as Star and Ruhleder argue by describing how infrastructures come to be learned as part of membership and conventions in a community of practice \citep{star1996steps}.
Likewise, the ways fairness becomes conceptualized and operationalized is related to what community of practice is grappling with the term.
As computer scientists begin to address questions about values---fairness, and others---we cannot be content to allow social science and humanities scholars simply cede the stage to engineers and lawyers or to have separate conversations within each discipline.
Responsibly advocating for the use of system design to advance values requires attending to its politics from a perspective that attends to all disciplines, including the potential shift in power as terms are subtly, or perhaps brazenly, redefined to suit particular epistemologies, methods, or preconceptions.
The emerging community of researchers studying technology and values is keenly attuned to the ability of values to recede when embedded in technological forms.
A shared analytic can, we propose, allow at least experts to continue to see and call attention to values as they are absorbed and baked into technological systems, or discarded for practical or political reasons.

\section{Methods}
While the lens of infrastructures is often used to analyze practices of the creation, maintenance, and repair of sociotechnical systems---often through ethnographic study (e.g., \cite{Steinhardt2014Reconciling,Steinhardt2016Breaking,Rosner2014Designing})---we take inspiration from Irani and Silberman's call for researchers to (re)design infrastructures as part of a constructive agenda \citep{Irani2014Critical}. 
The framing of our own work as infrastructuring put us in a more reflective frame of mind while exploring the literatures of our respective fields.

The authors are part of several overlapping communities focused on addressing values in socio-technical systems: three are part of a two year interdisciplinary effort at the University to specifically explore issues of fairness and opacity in algorithmic systems; two have collaborated extensively on research at the intersection of privacy and design; two are part of the broader multistakeholder effort to address fairness, accountability, transparency, and explainability in machine learning systems; and three regularly teach courses that explore values in socio-technical systems through a multidisciplinary lens. Collectively, the authors have observed and participated in a range of activities and venues addressing fairness in machine learning systems where they have witnessed disagreement and confusion caused by definitional and other forms of slippage, as fairness and other core concepts cross disciplinary boundaries. These experiences led us to understand the urgent need for infrastructure to support greater explication of the concept of fairness--and related terms--in interdisciplinary collaborations. Thus, the purpose of our investigation was to support interdisciplinary work within our own local communities and in the broader community of researchers and practitioners working on fairness in machine learning systems. 

We began this task by drawing on our own varied backgrounds, culling through literature discussing fairness (and related concepts of accountability, transparency, and explainability) in our own disciplines and fields--applied mathematics, computer science, data science, human computer interaction, law, statistics, and values in design. This starting point also aligned with our goal of providing an analytic for practices related to fairness, as these fields employ fairness in relation to some type of practice (e.g. building a system, going to trial, etc). Our research quickly snowballed to include core philosophical texts discussing fairness which were often referenced in our disciplines and fields. Our goal was to use this process to surface and reflect on the range of different meanings and usages of these terms across disciplinary groups, not to make a judgment about a ``correct'' meaning, nor to provide an exhaustive review of every paper discussing fairness. Thus we approached the literature review with a reflexive framing--documenting not just the naked definitions proffered but the underlying assumptions and commitments they reflect. This approach foregrounded the social and political implications of the divergent ways our respective communities use the term fairness, and highlighted the urgent need to put these definitions of fairness more directly in conversation and relate them to conceptions of justice. The approach also foregrounded different epistemological orientations towards fairness, such as viewing fairness as a mathematically achievable state, or part of continuous social justice practices, or being co-constituted by social and institutional forces. This research also drew our attention to fundamental issues essential to the collective research around values in systems such as the level at which we interrogate systems as reflected in the discussion of what different communities mean when they use the terms algorithm or system.

Through group discussion and iterative categorization and writing, we initially developed a definitional overview of concepts, and realized that supporting interdisciplinary inquiry required a tool that would allow the concept of fairness to be explored and debated through dimensions that cut across disciplines. Through our process of discussion and writing, the definitional overview of concepts evolved into an analytic to be used as a conceptual tool. Our decision to develop an analytic was based in part on our experience with the privacy analytic by Mulligan et al.~\cite{mulligan2016privacy}---a tool developed to support more nuanced discussions of competing conceptions of privacy, by providing a set of cross-disciplinary dimensions of privacy rather than a set of definitions or meanings of privacy. Through iterative group discussion, working through examples that emerged from the literature and our experiences, and reflection on our experiences in other interdisciplinary workshops tackling values in machine learning~\cite{fat,fatml,kroll-kohli-neurips-tutorial,afog}, we refined our understanding of what dimensions (and related questions) might ground interdisciplinary conversations and collaborations on fairness in machine learning. 

We view the fairness analytic as something that can be used throughout a values in design or ``values at play'' process~\cite{nissenbaum2005values,flanagan2005values}, but think that it may be particularly useful during initial exploratory phases of values discovery and identifying values-based conflicts by providing a way to think and discuss  across technical, philosophical, and empirical domain aspects of ``fairness.''

\section{``Conceptions of Fairness'' in Computer Systems}
\label{sec:taxonomy}
In an effort to build an infrastructure to support interdisciplinary research on building technology that supports human values, this section presents the fruits of our research into the use of ``fairness'' by different stakeholder communities.
When terms that cross cultural boundaries are not available, one alternative approach to facilitating better communication is to describe concepts as they exist in different disciplines.
Below we describe, and to some extent explore and contrast, how different groups understand, forecast, and approach the term ``fairness'' when it arises as part of the work of researchers and practitioners.
The goal is to describe, clarify, and sharpen differences in the use of the term as they appear across disciplines.
Our investigation revealed how the use of the term ``fairness'' itself varies, and in particular some of the dimensions along which it varies; we have organized these dimensions of variance into our fairness analytic (See Section~\ref{sec:analytic}).
 Sometimes we find concepts and super-concepts; sometimes the use of the term between disciplines overlaps but incompletely; and sometimes the term is used in different disciplines to address adjacent ideas rather than the same idea.
The variety of conceptions of fairness considered here exemplifies the need for an analytic to bring these diverse vocabularies into conversation.
We chose to focus our efforts on understandings of fairness both as it is the most studied concept in this emerging field and because it is the concept with the richest variety of conceptual approaches and which has generated the most strident disagreements as it stands to be realized in computer systems.
However, this work draws from a larger effort to understand the variety of conceptualizations of many terms relevant to and used in inquiries into values in computing, including ``privacy'', ``accountability'', ``transparency'', and ``explanation''.
Each of these terms, like fairness, evokes rich concepts understood differently in different fields, and each leads to conflict over the value of interventions depending on the disciplinary lens through which it is understood.

In describing the use of these terms by different communities, we also describe how those communities understand contrast concepts for those terms (the concept that properly contrasts to the specific conception of the term or which a user of a term intends to dispel by that use).
Contrast concepts negatively define the contours of the concept in question and are particularly useful in disambiguating abstract concepts that can sensibly be broken down positively in several different ways.
We also consider whether disagreement about the meaning or scope of other terms leads to downstream misunderstandings about the propriety or applicability of particular interventions (for an example, see the discussion of the use of the term ``algorithm'' in Section~\ref{sec:unit-of-analysis}).

Our discussion of terminology provides the opportunity to consider the frontier of research on the core topic of fairness as it is represented in this definitional overview, contrasting different approaches to its operationalization according to the views of different possible viewers and stakeholders.
For example, we discuss the forefront of research in machine learning fairness, viewed through the lens of how the term ``fairness'' is used by computer scientists, practicing lawyers, law scholars, social scientists, philosophers, and others.
This discussion allows us to suggest approaches to bridging the gaps between these constructs and highlights opportunities for each community to make use of the work being done by scholars and practitioners in other communities.

Fairness is a deep subject addressed by an enormous literature spanning thousands of years and many disciplines.
Philosophers, lawyers, and humanists of all stripes consider fairness in terms of morality and dignity, though also in terms of process and allocation.
Policies and laws designed to protect individuals and groups from discrimination or mistreatment tend to focus on proscribing behaviors, biases, and basing decisions on certain protected factors---in this last sense, antidiscrimination laws share a kinship with privacy laws, in that both seek to eliminate the spread or influence of certain sensitive information and serve to protect dignitary interests.
Social scientists often consider fairness in light of social relationships, power dynamics, institutions, and markets.
Here, there is an emphasis on the constructed nature both of perceptions of fairness and on the factors which might affect unfairness or discrimination.
Finally, quantitative fields have studied questions of fair allocation as pure mathematical problems, through the design of market or voting mechanisms for eliciting outcomes that appear fair, and in the sorts of statistical and machine learning models which have triggered renewed interest in the values implications of software technology.

\subsubsection{The Philosophy of Fairness}
Ideas about fairness often rest intuitively on the sense that what is fair is also what is morally right, especially as something which is not morally right is likely also not fair for some actor.
We provide here a brief and necessarily incomplete overview of the philosophical underpinnings of these ideas.
A more detailed treatment, not undertaken here, would relate the conceptions of fairness articulated here to ideas about operationalizing fairness in specific ways in technical systems, as we describe below.

Theories of morality and ethics are often broken down by whether they consider a need for ``doing the right thing'' to stem from virtues, duties and rules, or consequences.
\emph{Virtue ethics} is the ancient discipline of moral analysis based on virtues---deep-seated traits of character which make the person who has them morally good.
In the frame of virtue ethics, fairness is simply a virtue, to be perfected by practice and learning, and embodied in a moral agent.
This notion of fairness undergirds the idea that properly designed technologies can be sufficiently virtuously behaved as to be fair in a given application.
Nussbaum examines virtue ethics as presented by the Greek philosophers, finding that a focus on virtues leaves open the possibility that external factors can compromise virtues and suggesting reason as a necessary component of realizing virtue in human flourishing~\cite{nussbaum1986fragility}.
She has also questioned whether virtues constitute a separate category for the basis of ethics than duties~\cite{nussbaum1999virtue}.

Much of \emph{deontological ethics}---that is, ethics based on the duty to uphold given rules---passes through the work of Immanuel Kant to modern philosophers~\cite{korsgaard1996creating, korsgaard1996sources}.
Kant argues for upholding the \emph{categorical imperative}, a rule which defines behavior as moral if it is consistent with a world in which everyone takes up that behavior.
Fairness derives from duty in this frame, attaching moral valence to an action by demanding that what fairness requires is embodied in a duty, and what is unfair constitutes a breach of this duty.
This notion of fairness is a basis for legal notions of fairness, which often proscribe specific types of unfairness or assign a duty to uphold certain requirements or to perform certain actions to a specific actor.
Schopenhauer critiques this reliance on duty, however, rooting morality in the virtue of compassion rather than in the cold and dead rote application of a rule~\cite{schopenhauer1840basis}.

Finally, \emph{consequentialism} defines morality in terms of the consequences of certain actions~\cite{sinnott2003consequentialism}.
The most straightforward operationalization of this comes in the form of \emph{utilitarianism}, an approach to assigning moral valence to actions straightforwardly by calculating the total amount of good and bad in the consequences of those actions~\cite{bentham1789introduction, mill1861utilitarianism}.
The net good---the total amount of good (for all people) minus the total amount of bad (for all people)---for a morally right action must then be greater than the net good for any other available action.
Classic utilitarianism differs from deontological ideas about morality because it denies that anything matters in assessing the morality of an action other than the consequences.
Fairness derives from the setting of the utilitarian view: if an action maximizes the overall good, it must of necessity be the most fair action, since it would be unfair to someone to take another action (because in that scenario, that person would receive less good and would thus be treated unfairly).
Utilitarianism underlies many technological operationalizations of fairness, which function by counting up the good (say, correct outputs of a software scoring system) and balancing it in some way against the bad (say, erroneous outputs of that same system).
For an example of this and how it trades off with other ideas about fairness, see Section~\ref{sec:risk-assessments}.

Political philosophy has also engaged questions of fairness as a matter of social and societal organization.
Rawls, for example, defines notions of distributive justice and equity as fair when they are independent of one's place in society, subject to the ``difference principle'' that deviations from strict equality are acceptable to the extent they make the least advantaged better off than they would be in a world of strict equality~\cite{rawls2009theory}.
Rawls refines this framing to the idea of ``justice as fairness''~\cite{rawls2001justice}.
Therefore, the only societies which are fair are those we would choose to live in (or contract for with others) absent knowledge of our position within them, restricting analysis to behind a ``veil of ignorance''.

Another approach articulates the concept of ``luck egalitarianism'', the idea that factors which are left up to chance should not (in a normative sense) interfere with one's access to a fair set of opportunities in society.
That is, a fair society is one in which outcomes are based on the responsible choices people make, not on uncontrolled aspects of their circumstances.
While this theory is not in conflict with Rawls' justice as fairness~\cite{tomlin2012can}, it has been critiqued by Anderson, who notes that the goal of political fairness should be ending socially imposed oppression, not eliminating the effects of ``brute luck''~\cite{anderson1999point}.
Many software systems are designed to treat all subjects in procedurally equal ways, so the question of how the luck of ability and situation of those subjects relates to that treatment is relevant to interrogations of the fairness of the technical artifact.

Other philosophers have viewed fairness in the context of notions of general agreement, viewing right and wrong as subject to agreement at the societal level or at the individual level (possibly subject to assent for that agreement at the societal level).
These \emph{contractual} notions of right and wrong bear some unpacking for our purposes, as they vest the question of what is moral and fair in agreements among affected and unaffected people.
Social contract theory has its roots in Hobbes, who begins with the fundamental self-interest of individuals~\cite{hobbes1651leviathan} and argues that it is most beneficial for this self interest to form agreements about moral norms.
Government power, then, and the normative force of laws and morals, both stem from the consent of those agreeing to be bound by them, who do so because it is in their interest to make such arrangements.
Rousseau~\cite{rousseau1762social}, also argues that the sovereign power of society stems from the willingness of its participants to give up natural freedoms for equally burdensome laws and duties.
However, Rousseau argues that this agreement stems from the fact that rationality requires respect for persons, and that mutual respect must form the basis for the contents of the social contract.
Ideas about social contracts are often subdivided along these lines, with ideas grounded in Hobbes' view falling under the banner of \emph{contractarianism} and ideas stemming from Rousseau under the banner of \emph{contractualism}.
Scanlon, a key modern contractualist, explicitly relates fairness to the construction of social contracts, noting that we can consider fairness in terms of the effects it has on others we interact with~\cite{scanlon1998we}.
He also argues that human rights matter only insofar as the consequences of having those rights can be articulated and experienced~\cite{scanlon2004does}.
Social contract theory helps when analyzing values in software systems, as it speaks to what is necessary to make outcomes mediated by those systems legitimate embodiments of those values.
The language of contractarianism also arises often in discussions of the purported benefits and costs of technology adoption, which is based on the interests of those involved as weighed against the interests of others.

Another approach to breaking down ideas of fairness in philosophy is based on the operative concept of justice served by that moral theory.
Many of the key debates about what fairness means or what it entails happen within these frames, and several distinctions apply~\cite{miller2017justice}.
Rawls and Scanlon both view fairness through the lens of distributed substantive justice,
focusing on how to allocate resources in a society or sphere of social or political life.Those allocations may vary (and the variance be ascribed moral rectitude) based on whether and to what extent attributes of individuals (both within their control and outside it) should inform distributions; whether need, or effort, or merit, or societal contribution should serve as the basis for distribution; and even what is to be distributed resources or opportunities.
Justice and fairness in this framing both serve to respect and to support human dignity.
Others operationalize justice in a procedural mode rather than a substantive one, focusing on the extent to which procedures apply equally to all, or consider justice as corrective, applying remedially when one person interferes with another's allocations rather than as a principle for understanding the correct allocation of distributable goods and positions within society.
Beitz, for example, understands political fairness as a ``normative property of institutions and procedures''~\cite{beitz1989political}, though his view considers fairness at the level of political agency in shaping those institutions and procedures rather than requiring that the institutions and procedures themselves be fair.
Broome describes fairness as embodied in the process of a lottery, explaining why lottery allocation matches intuitive ideas about fairness~\cite{broome1990fairness}.
Ryan rejects the notion of contractualized fairness and fairness as a property of institutions and procedures in summarizing several schools of thought on fairness in political philosophy~\cite{ryan2006fairness}.
Hellman draws distinctions between fairness and justice in understanding the underpinnings of why discrimination is a moral harm~\cite{hellman2008discrimination}.

Legal philosophers are more engaged with the question of how to embody fairness in law.
Hart's foundational positivist work frames laws as rules constructed by humans, without regard to whether those rules in fact capture morality~\cite{hart1961concept}.
Dworkin, by contrast, argues that human rights exist outside the written law and that the law can be superceded by such interests~\cite{dworkin2013taking}.
Many of the same debates take place when considering fairness as embodied in a technical artifact: to what extent are the rules constructed of the artifact itself vs. deriving from an absolute notion of right and wrong?

\subsubsection{Unfairness and Discrimination}
Understandings of fairness can be further interrogating through exploration of \textit{contrast concepts}, or concepts that are placed in opposition to fairness \cite{mulligan2016privacy}.
By examining what is explicitly seen as not fair, we can better understand what ``fairness'' is deployed to achieve and better understand its scope and limits.
Contrast concepts to fairness include \textit{unfairness} and \textit{discrimination}.
In law, there are frameworks via equal protection laws to address and minimize unfairness and discrimination.
Disputes over the concept of equal protection illustrate divergent perspectives on fairness.
Does equality before the law demand recognition of existing racism and allow actions to remedy it, or demand "color blindness"? 
For example, compare, the majority opinion in \emph{Grutter v. Bollinger},\footnote{539 U.S. 306, 327--41 (2003)} which upholds, under a strict scrutiny analysis, affirmative action in law school admissions) with the opinion of Justice Thomas, concurring in part and dissenting in part, which argues that the equal protection clause prohibits such affirmative action because it classifies on the basis of race.

U.S. equal protection laws vary by sector, including the Fair Housing Act (prohibiting discrimination in housing decisions), the Americans with Disabilities Acts (prohibiting discrimination on the basis of covered disabilities), or Title VII of the Civil Rights Act and the Age Discrimination in Employment Act (prohibiting employment discrimination) but share a focus on limiting the impact of protected characteristics--race, color, religion, sex, national origin, disability--on significant life opportunities. 

``Disparate Impact'' and ``Disparate Treatment'' are considered forms of unfair discrimination; to prove that one of these equal protection laws has been violated, plaintiffs must show that they have faced one of these forms of discrimination. 

\textit{Disparate treatment} occurs ``when people are treated differently on the basis of a protected characteristic''  \citep{Bodea2018EUautomated}. In U.S. law, these protected characteristics include race, color, sex or gender, religion, age, disability status, national origin, and marital status. For instance, a decision about providing credit to someone cannot be made because of one's marital status---even if there were data that suggested that marital status is correlated with one's credit score.

\textit{Disparate impact} ``occurs when a company employs facially neutral policies or practices that have a disproportionate adverse effect or impact on a protected class, unless those practices or policies further a legitimate business need that cannot reasonably be achieved by means that have less disparate an impact.'' \cite{FTC2016BigDataTool}. 

For instance, making decisions about credit decisions based on consumers' ZIP codes could potentially have a disparate impact on an ethnic group, as some ethnic groups are concentrated in particular ZIP codes and geographic areas \cite{FTC2016BigDataTool}. In both cases, there are particular legally protected characteristics or protected classes that the law has identified as needing specific protections from unfair discrimination in various domains, including housing, employment, and access to financial credit. 

\textit{Due Process}
In the context of decision making about individuals the term fairness also applies to process. Due process imposes procedural constraints on government actions that are measured by fairness, as well as risk of erroneous deprivation, the seriousness of those risks, and the costs of providing more process.\footnote{Mathews v. Eldridge 424 U.S. 319 (1976)} For decision making processes to be considered fair they must support participation by subjects--for example providing them with access to the data and logic being used to render decisions, as well as opportunities to correct or challenge them--and additionally protect dignitary interests that provide for a voice in legal proceedings or require human oversight over technological processes. When protected interests are involved hearings which provide opportunities for an individual to participate, understand evidence, and shape outcomes are an important component of fairness. This concept of fairness is captured in data protection law as well, such as in the European General Data Protection Regulation.
Both data protection and general due process attend to fair decision-making processes. 

\subsubsection{Fairness and Society}
Questions of fairness are often viewed empirically, situated within society.
Such questions are the general purview of social scientists, who have developed relevant approaches to analyzing fairness and discrimination.
For example, while decision policies can be disparate in their direct treatment of individuals or inadvertently in their impact, neither fully reaches the idea that the opportunities available to individuals may be constructed of their position in society~\cite{pincus1996discrimination}.
Further, critical scholars of race have noted the tendency of advantages to accrue to members of certain groups under the heading of \emph{privilege}~\cite{mcintosh2001white, Leonardo2004}.
In particular, if one subscribes to a constructivist theory of the attributes that define membership in privileged and underprivileged groups, it seems inherently problematic to consider these factors as ``attributes''~\cite{kohler2018eddie} Similarly the concept of \emph{intersectionality}, which identifies the ways that race, class and gender are mutually constitutive and interact in lived experiences of oppression \cite{crenshaw1990mapping}, is at odds with such an atomistic approach.  Race, as a social construct cannot be viewed as a bit or variable that can be removed or controlled in an analysis, but rather shifts the gaze to the myriad ways in which racism is reflected in data collection and other social practices and the systems of classification that use and support them.

\subsubsection{Quantitative Notions of Fairness}
In addition to qualitative aspects deriving from ideas about morality and ethics, fairness as a matter of equal or equitable allocation has a long history of study in mathematical fields. Technical definitions of fairness are manifold, and can be substantially different to one another. Within quantitative fields such as mathematics, computer science, statistics, and economics, fairness criteria are generally defined with respect to a specific task or problem. We explore three problems where quantitative approaches to fairness have been considered: fair division, voting, and fair machine learning.\footnote{This leaves aside other interesting problems of fair resource allocation in computer science (e.g. process scheduling, network throughput, disk allocation) where ``fair'' generally means ``in equal portions'' or ``in proportion to the amount requested'' or ``available within a reasonable amount of time''. In these problems, interesting distinctions come via questions of the unit of analysis: are equal resources provided to low-level primitives (e.g., processes, network flows) or to particular logical abstractions (users, groups, network endpoints)? That is not to say that there are not rich questions of fairness to be had even in this simpler arena; network traffic discrimination and the allocation of bandwidth are often the subject of both contractual and political constraints (for example, under the banner of ``network neutrality''~\citep{wu2003network}).}

\paragraph{Fair Division}
The simplest fair allocation problem is that of \emph{fair division}, or ``cake cutting'': if one has some resource (``cake''), how would one best allocate that resource (``cut'') according to some definition of \emph{utility}, the benefit accrued from each allocation?
In an interactive setting, each party is incentivized to maximize a fair version of utility in a ``cut-and-choose'' protocol, where one party proposes an allocation and another proposes the assignment of shares to parties.
Such allocations satisfy the condition that they are \emph{envy-free}, meaning that no party should prefer the shares given to another party, on the basis of the separation of allocation and assignment.
Fair division protocols can have other goals, such as \emph{proportionality} (the property that each participant receives a share at least as large as their fraction of the parties) and \emph{equitablilty} (the property that each participant derives equal utility from their share)~\cite{procaccia2015cake}.

Lee and Baykal observe that ``optimally fair'' allocations determined mathematically based on expressions of utility may not be preferable to negotiated allocations for various reasons~\cite{lee2017algorithmic}.
This suggests that people do a poor job expressing their utility mathematically or that the dignitary interest served by negotiation strengthens the perceived fairness of outcomes.
This suggests that mathematical notions of fairness in the allocation of resources may be insufficient on their own to achieve fairness (or even the perception of fairness) in sociotechnical systems.

\paragraph{Voting}
Fairness has also received much attention in the voting literature, a part of game theory in which voting is a mechanism where every participant's preferences are collected (in \emph{ballots}) and then those preferences are combined into a \emph{outcome}.
Outcomes can take many forms: a single winner, a collection of winners, or a socially-aggregated ranked list, or even no winner.
Various intuitive notions of fairness cannot be achieved by many voting mechanisms, as shown through the plethora of impossibility results in this space for both deterministic \cite{arrow1950difficulty, gibbard1973manipulation, satterthwaite1975strategy} and randomized \cite{kohli2018epsilon} schemes.
Such desiderata include equality (the votes of all voters are treated the same), neutrality (all candidates on a ballot are treated the same), monotonicity (if candidate A is already the winner, a hypothetical transfer of votes from B to A maintains A as the winner), unanimity (if everyone prefers candidate A over B, then the voting system should rank A above B as well), non-dictatorial (the voting system is not responsive to the vote of a single individual in all cases), and non-imposition (a path exists for every candidate to win).
Sometimes, changes in the setting or assumptions can circumvent these impossibility results \cite{may1952set, moulin1980strategy}.
In general, however, voting as a pure mechanism also suffers in terms of its ability to capture broader ideas about fairness.
  
\paragraph{Fair Machine Learning}
Recent attention to the use of machine learning in socially critical applications has focused research energies on the problems of fairness in machine learning and statistical models more generally~\cite{hardt2018fairml}. Machine learning researchers have operationalized demands that systems treat people fairly with respect to \emph{sensitive attributes} (e.g., gender, race, etc.) in many ways, grouped broadly into two categories: \emph{individual fairness} (i.e., treating like cases alike)~\cite{dwork2012fairness}; and \emph{group fairness} (i.e., treating groups---often defined by legally protected attributes---in a way that is somehow equal or equitable, such as by matching various statistics across groups)~\cite{mitchell2018mirror, narayanan2018twentyone, mitchell2018catalogue, hardt2016equality}. Both ideas stand in contrast to an intuitive notion of \emph{fairness through blindness}, the idea that simply removing attributes on the basis of which one does not wish to make predictions from the data upon which models are built will eliminate its effect from the model. Blindness fails for many reasons, notably because it does not in fact, on its own, remove the impact of the data being blinded, which may be \emph{redundantly encoded} in other factors. As such, Lipton, Chouldechova, and McAuley have suggested that fixing disparities in impact may inherently require creating disparities in treatment~\cite{lipton2017does}. Indeed, in cases where the attribute one desires to suppress is socially constructed, it may not be meaningful without this proxy information~\cite{kohler2018eddie}.

Individual fairness requires similar people are treated similarly by the classification mechanism. But this approach requires knowing how to measure how ``similar'' two individuals are, at least for the purpose of the particular task to which machine learning is being deployed. Further, the fact that similar individuals are treated similarly does not address the question of whether that treatment is \emph{equitable} or \emph{justified}, both key components of fairness as conceptualized philosophically. 

Group fairness can be formulated in many ways as well. Perhaps the simplest group fairness criterion to state and the most natural measure for which to strive is \emph{demographic parity}, the property that the acceptance rate of an algorithm (that is, the fraction of inputs which receive a positive classification) should be the same for subgroups across all values of a particular sensitive attribute (e.g., the same fraction of white and black internet users see a particular ad). Mathematically, this is equivalent to the outcome being statistically independent from the sensitive attribute (i.e. the chance of seeing the ad does not depend on race). On the surface, demographic parity is attractive because it describes a world in which decisions truly do not depend on attributes considered sensitive.

Not all statistical fairness criteria are created equal, and none fully captures the richness with which fairness is conceptualized in other fields. If proper care is not taken in the operationalization of such criteria, systems may still exhibit undesired behavior~\citep{hardt2018fairml}. For example, although demographic parity seems innocuous and even a desireable criterion on the surface, it suffers from serious design flaws. Since demographic parity only asks that the acceptance rate across subgroups be the same, without specifying how this be achieved, we could satisfy this requirement in a perverse manner. Suppose a certain model has an 80\% acceptance rate for men. We can enforce demographic parity as follows: for women, simply pick a number from 1---10 uniformly at random; if the value is 9 or 10, we reject the woman; otherwise, we accept them. In this case, the acceptance rate for women matches that of men at 80\%. However, the mechanism does not use any of the subject's characteristics to make its decision. Though we have achieved demographic parity, we have created a situation that seems unfair in other ways---applying a different and arbitrary rule for women vs. men.

Even if we demand that the rules for each group be sensible, demographic parity fails to capture information about which members of a group should receive any given outcome (so, in the advertising example above, a system could show the ad only to white users who are interested in or qualified for the product or service advertised and only to minority users who are not interested or qualified or who simply cannot afford the good or service on offer---so an ad ``shown equally'' across groups may still lead to disparate real-world outcomes because members of each group respond to it differently). Still, demographic parity remains a seductively straightforward demand for fairness. It is likely better described as an ideal: in a fairer world where opportunities in fact did not depend on protected status attributes, demographic parity would occur naturally. However, given that opportunities are distributed disparately and unequally in the real world, such a demand is generally over-restrictive for real decision systems while still allowing for significant unfairness.

To capture more expressive ideas of fairness in statistical definitions, other criteria have been proposed based on notions of statistical parity: explicitly demanding equalized error rates amongst groups, at the cost of further constraining model behavior and possibly reducing overall accuracy. Other proposed metrics include \textit{equalized odds}~\citep{hardt2016equality}, equalizing the chance that a member of each group be accepted conditioned on that member's true condition (e.g., selecting the same fraction of black borrowers and white borrowers who will actually pay back a loan) and \textit{predictive value parity}~\citep{chouldechova2017fair}, which equalizes positive or negative predictive value, a statistic which represents the chance that positive or negative predictions are in fact accurate (that is, the predictive value of the classifier for membership in the positive or negative class).\footnote{For an overview of many available statistical fairness measures, see several tutorials at major machine learning research conferences~\citep{bonchi2016bias, hardt2017fairness,narayanan2018twentyone} as well as excellent summaries in~\citep{mitchell2018mirror, mitchell2018catalogue, verma2018fairness}.
Finally, an online book-in-progress presents a more complete case around measures of discrimination~\citep{hardt2018fairml}.} 

Somewhat pessimistically, statistical fairness research is riddled with impossibility results. That is, in the context of a particular task, it is in general impossible to construct an algorithm that satisfies all of the above-described fairness criteria~\citep{kleinberg2016inherent,chouldechova2017fair}. Conceptually speaking, something has to give: considerations of fairness inherently pit different equities against each other and fairness is often described as essentially contested~\citep{gallie1955essentially}. Purely mathematical intuition suggests that placing additional restrictions on the behavior of an algorithm for a particular use can only limit the set of remaining acceptable algorithms for that application. Demanding that an algorithm to satisfy too many fairness criteria can restrict the class of potential algorithms so much that it implies no solution to the problem at hand. 

Stepping away from mathematical technicalities, Jacobs and Wallach have argued that issues in fair machine learning come from a more fundamental source, namely the failure to attend to issues of measurement and construct validity in models~\cite{jacobs2018measurement}. \emph{Construct validity} is a literature from statistics that evaluates models not only based on their predictive capacity but on their structure~\cite{messick1987validity, quinn2010analyze}.

Alternatively, Kusner et al. have argued that fairness in machine learning model outputs can be understood counterfactually in terms of how the inputs can be modulated to achieve different outcomes~\cite{kusner2017counterfactual}. This leads into the idea of strategically manipulating one's attributes in order to get the most desirable output from a model and the related question of how to build a model to resist such gaming~\cite{hardt2016strategic}.
Notably, strategic behavior may have disparate effects (as individuals may have disparate access to the instruments of manipulation)~\cite{hu2018disparate} or broader social losses~\cite{milli2018social}. Economists have long noted that such strategic behavior can cause statistical relationships to break down and be arbitraged away through the phenomenon of \emph{Goodhart's law}~\cite{goodhart1984problems}.

In our discussion of quantitative notions of fairness, the fairness criteria proposed are more akin to specific measures of types of \emph{unfairness} one would hope a reasonable system does not exhibit than they are to holistic concepts of fairness. Indeed, it is perhaps best to think of such criteria not as specifications of fairness, but as operationalizations of criteria it may be unreasonable to violate. This is underscored by mathematical impossibility results and the observation that fairness is essentially contested, both of which imply the necessity of tradeoffs and the lack of any truly neutral design choices. Mathematical fairness criteria should not be confused with sufficient conditions for fairness---these criteria aim to protect a particular harm or injustice. Indeed, impossibility results tell us that some of these criteria must be violated in any interesting application. In particular cases, however, certain of these conditions may be necessary to achieve fairness; certainly, they provide useful and technologically sound approaches to breaking down fairness problems---the question is when and how to apply which ones.

\section{Fairness Analytic}
The wide ranging meanings of fairness detailed above presents challenges to collaboration, but also presents opportunities for deeper exploration of what fairness could and should mean in various sociotechnical systems. Working productively amidst the wealth of definitions however requires tools to bring them into conversation and enable researchers from different disciplines to work fluidly across and among them. This counsels against rigid definitions and towards a tool that allows researchers to reflexively explore various conceptions of fairness both within and across communities. For these reasons, we chose to develop a conceptual analytic that highlights specific dimensions along which various definitions of fairness described above can be considered. The Fairness Analytic allows researchers to bring the definitions into conversation from different vantage points. 

The Fairness Analytic can be found in Appendix A and contains six overarching dimensions: theory, relationship to society, unfairness, protection, provision, and context along which conceptions of fairness can be unpacked. Our claim is that analytically separating these overarching dimensions can clarify the function and value of fairness in practice. Dimensions of theory help focus collaborators on the purpose and justifications of fairness, politically important yet often underspecified in discussions and arguments about the design of computer systems. Relationship to society supports clarification of important connections between concepts of fairness and justice, as well as the bounds of the system or sociotechnical system under scrutiny. Dimensions of context further bound fairness to domains, time-frames, and ideas about measurement. Dimensions of unfairness supports elaboration on what and who is considered capable of violation fairness, often significant considerations in the design of sociotechnical systems. Finally, the dimensions of protection and provision help clarify both more specifically what fairness is sought to provide, whom it is to protect, and where such protection can be operationalized.  

Below we describe and explore the dimensions and sub-dimensions more fully through application to a case study. 


\section{Case Study: Debating the Fairness of Criminal Risk Assessments}
\label{sec:risk-assessments}
The community that works with values in computer systems is familiar with the challenges posed by communicating across disciplines.
A shared infrastructure, including but not limited to a shared framework (such as that provided by our analytic) for describing, comparing, and contesting values such as fairness, is particularly important to support the analysis and design of real world systems in settings with significant values-oriented issues. 

A sharp example of where a framework for organizing discussion across stakeholder groups and disciplinary approaches around fairness in the application of computer systems in the real world comes from public debates about the recidivism risk assessment systems---tools used to predict an individual's risk of engaging in future crimes, proxied by their likelihood of re-arrest, in the context of criminal justice decisions such as those concerning arrest, bail, parole, or sentencing.
Debates about fairness within such systems clearly demonstrate the confusion stemming from different conceptions of ``fairness''.
Legislatures have moved aggressively to require the use of risk scoring in some contexts~\citep{hlr2017loomis,pai2019riskassessments}, particularly at sentencing as an effort to improve fairness and the overall management of criminal justice.\footnote{One effort, California's SB 10, pairs the criminal justice reform goal of eliminating money bail with a requirement for procuring and using risk assessment tools, but its implementation has been stalled by a recall of the law.}
From their perspective risk assessment tools promise to provide an ``objective'' assessment of an individual's criminal risk.
Concerned with the unfairness that results from human bias, legislatures have sought to use these tools because---as algorithms---they don't suffer from latent human biases that plague judges, such as outgroup bias, availability heuristics, confirmation bias, and other cognitive biases and phenomena.
They viewed this as a move to increase the fairness of bail or jail outcomes by reducing a known source of unequal treatment based on perceptions of the accused's race.
Many communities (including the research and advocacy communities) have, however, adopted a range of perspectives from seeing the tools as a positive addition to the criminal justice toolkit~\cite{berk2019machine, flores2016false}, to tools that trade off different ideas about fairness that therefore require careful governance~\cite{berk2018fairness,pai2019riskassessments}, to risky tools with the potential to mislead judges~\cite{green2019disparate} or overcomplexify simple relationships~\cite{dressel2018accuracy}.

The fairness of the criminal recidivism risk assessment tool COMPAS (developed by Northpointe, Inc.), as applied to the assessment of pre-trial sanctions, has been the subject of heated public debate, litigation, and discussion and research in several communities.\footnote{The use of COMPAS in other contexts, such as at sentencing, has also been subject to much of the same technical criticism and also litigation in multiple states~\cite{hlr2017loomis}.}
Much ink has been spilt detailing and arguing about the system (e.g., \citep{propublica-bias,dieterich2016compas,flores2016false,kleinberg2016inherent,chouldechova2017fair,corbett-davies2016computer,corbett2017algorithmic,berk2018fairness,dressel2018accuracy,pai2019riskassessments,green2019disparate,goel2019accuracy}).
Our discussion here is limited, aimed at showing the way in which different, and often under-specified, conceptions of ``fairness''---along with variations in how ``the system'' was defined---complicated the conversation, and limited the generative potential of the disagreement, for example to clarify areas of consensus and disagreement about facts, assumptions, and desirable end states.
Informed by our exploration of conceptions of fairness, organized into the fairness analytic, below we explore the COMPAS debates to highlight the potential benefit of cross-disciplinary understanding.

\subsection{``Algorithms'' and the unit of analysis}
\label{sec:unit-of-analysis}
Determining the appropriate unit of analysis or level of abstraction at which to engage questions about COMPAS is its own challenge.
We give a brief aside about the meaning of the seemingly uncontested term ``algorithm'', in order to highlight the depth at which terminological confusion and lack of a shared vocabulary governs discussion of real cases.
As with the more abstract term fairness we consider in Section~\ref{sec:taxonomy}, the term ``algorithm'' is assigned disparate meaning in the literatures of computer science and other fields, and is even contested within computer science.
For example, Donald Knuth defined algorithms as separate from mathematical formulae in that they must: (i.) ``always terminate after a finite number of steps'', (ii.) that ``each step must be precisely defined; the actions to be carried out must be rigorously and unambiguously specified for each case'', (iii.) that input is ``quantities which are given to it initially before the algorithm begins'', (iv.) that output is ``quantities which have a specified relation to the inputs'', and (v.) that ``all of the operations to be performed in the algorithm must be sufficiently basic that they can in principle be done exactly and in a finite length of time by a man using paper and pencil''~\citep{knuth1968art}.
Similarly and more simply, in their widely used textbook, Cormen, Leiserson, Rivest and Stein define an algorithm as ``any well-defined computational procedure that takes some value, or set of values, as input and produces some value, or set of values as output''~\citep{CLRS}.
Algorithms may be fully deterministic, mapping inputs to outputs as a mathematical function, or they may be randomized, mapping inputs to probability distributions of possible outputs, meaning the values may relate to the inputs only probabilistically.
These definitions are precise, clear, and scoped only to technical artifacts, though they differ in specificity somewhat.

By contrast, communications scholar Christian Sandvig says that ```algorithm' refers to the overall process'' by which some human actor uses a computer to do something, including decisions taken by humans as to what the computer should do, choices made during implementation, and even choices about how algorithms are represented and marketed to the public~\citep{sandvig2015seeing}.
Sandvig argues that even algorithms as simple as sorting ``have their own public relations'' and are inherently human in their decisions.
Another communications scholar, Nicholas Diakopoulos, defines algorithms in the narrow sense (``as a series of steps undertaken in order to solve a particular problem or accomplish a defined outcome'') but considers them in the broad sense (saying ``algorithms can arguably make mistakes and operate with biases'', which does not make sense for the narrower technical definition)~\citep{diakopoulos2015algorithmic}.

Science and Technology Studies scholar Malte Ziewitz asks whether we can examine algorithms ``as a figure that is mobilized by both practitioners and analysts?''~\citep{ziewitz2017not}.
Similarly, anthropologist Nick Seaver argues that algorithms are cultural objects situated in sociotechnical systems, constructed as much socially as technically~\citep{seaver2017algorithms}.

Technologists prefer to think only of the characteristics of their well-defined tools, scoped to and embodied within machines; by contrast, fields outside of computer science use the term ``algorithm'' to mean a full sociotechnical system that includes people and culture, not only a technical artifact.
There is an important gap in understanding between these views: if we wish to interrogate whether the application of an algorithm is ``fair'', we must first understand if we are considering the fairness only of the technical tool itself or if our analysis must also cover the tool's context and interactions with people and society. Even if considering only the technical tool the competing conceptions of algorithms discussed above presents challenges communication and collaboration across disciplines.

Clearly defining the unit of analysis sets the terms of the debate and also structures the nature of research questions.
For example, unfairness is often situated by technologists outside mathematical or technical artifacts but rather in the data fed into them or some other aspect of how they interact with society, preserving a notion of ``technological neutrality'' by rescoping the analysis.
Yet decisions about when and how to deploy which pieces of technology in which ways themselves have politics and can be arguably judged as more or less fair.

\subsection{COMPAS: An unfair algorithm?}
In May of 2016, the investigative journalism organization ProPublica released the results of their study of the criminal recidivism risk assessment tool, COMPAS, as it is deployed and used in Broward County, Florida~\citep{propublica-bias}.
The study revealed a startling bias in the tool's false-positive rates: black arrestees who would not be re-arrested in a two-year horizon were nonetheless scored into a ``high-risk'' category at nearly twice the rate of white arrestees who were not subsequently arrested~\citep{propublica-bias}.
The journalists argued that this bias made the use of COMPAS unfair to black people.
This makes intuitive sense as it appeals to fairness properties such as treating likes alike, and objecting to rules that, while facially neutral, produce disparate impacts along protected class status.
ProPublica claimed that assigning extra risk to certain people based only on their race was unfair.

The assessment's vendor, Northpointe (now, equivant), responded by asserting that COMPAS was, in fact, fair~\cite{dieterich2016compas}, as the scores are designed to define relative levels of risk (such that any score, say ``4 out of 10'', corresponds to the same risk of re-arrest regardless of race).
This also appeals to intuitive concepts of fairness in government decision-making, such as treating all citizens according to the same rules and demanding that bureaucratic assessments be accurate representations of the world.
After all, because the rate at which blacks are arrested in Broward County is nearly twice as high as the rate at which whites are arrested~\citep{propublica-bias}, satisfying the fairness criterion suggested in the ProPublica study would require that scores have different interpretations for whites and blacks~\cite{chouldechova2017fair}.

In a world where black people are arrested more often, a score that predicts the risk of arrest must be higher for black people. 
If it were not, then the same score would imply a different risk level for similarly situated white and black arrestees.
To remedy this, either the scores could be altered such that black people would receive lower scores than whites with similar attributes in some cases (thereby treating similar cases \emph{differently}), or the cutoff points could be altered, so that the perceived risk associated to a particular score would differ by race (for example, by setting the cutoff for a ``high risk'' designation higher for blacks than for whites).
Academics later pointed out that the demands of Northpointe and ProPublica---that scores correspond equally to risk for both whites and blacks, and that scores falsely create high-risk designations for whites and blacks at equal rates, which both seem necessary for the system to be fair---are mutually incompatible at a mathematical level~\citep{corbett-davies2016computer,kleinberg2016inherent,chouldechova2017fair}.
That is, ProPublica and Northpointe disagreed not on the performance of the tool or even the correct construction of the tool, but rather on what ``fairness'' requires in criminal justice applications.
This is not to say that either realization of ``fairness'' is correct or even to compare their moral weight; rather, each approach to operationalizing fairness in this context serves a particular stakeholder's interests.

All of these arguments considered the fairness of the system at a particular level of abstraction.
They focused on the inputs, outputs, and structure of the COMPAS tool itself, rather than its integration into the system of making pretrial decisions about bail and release, sentencing, or correctional offender management.
Yet the mathematical incompatability of operationalizations of fairness in this setting suggests that analysis might better be directed at the broader process.
The criminal justice system has an important due process interest in protecting the calibration of scores to measured re-arrest risk, so that scores can be interpreted by judges effectively~\cite{green2019disparate}.
Arrestees, on the other hand, have a strong interest in minimizing the false-positive rate for high-risk ratings, as well as in equalizing that rate across demographics~\cite{corbett2018measure}.
For both groups, however, the goal of reliably administering justice and minimizing the risk of failure to appear could be served instead by other behavioral interventions, such as messages reminding people of their court dates~\cite{cooke2018using}.

\subsection{Human vs. Machine Bias}
The turn to recidivism risk prediction was contested on the basis of the purported fairness of the algorithm itself, yet its adoption was driven by concerns with the fairness of the system at a higher level of abstraction. Given research showing that judges decision making can be affected by inappropriate factors~\citep{rachlinski2008does}, some suggest risk assessment as a tool to limit such bias.
Yet, as described above, the data used to create these risk assessments is historically situated within racially biased policing practices, and may not reflect true risk levels.
However, even setting aside concerns about biases in the data, how such tools interact with existing work practices is unclear and an area of active research~\cite{stevenson2018assessing, garrett2018judging}.
Ethnographic research by Angele Christin suggests that like other tools that are introduced into the workplace, the introduction of risk recidivism tools has met with resistance~\citep{christin2017algorithms}.
Her work suggests that at least part of that resistance is based on ideas about fairness.
Christin shares a quote from a senior judge she interviewed about analytic tools:

\begin{quote}
\begin{singlespace}
I don't look at the numbers.
There are things you can't quantify $[\ldots]$ You can take the same case, with the same defendant, the same criminal record, the same judge, the same attorney, the same prosecutor, and get two different decisions in different courts.
Or you can take the same case, with the same defendant, the same judge, etc., at a two-week interval and have completely different decision.
Is that justice? I think it is.
\end{singlespace}
\end{quote}

Similarly, Christin found that probation officers engaged in ``criteria tinkering.''  They would ``manipulate the variables they entered in risk-assessment tools in order to obtain the score that they thought was adequate for a given defendant''~\citep{christin2017algorithms}.
 Finally some of Christin's interview data suggests that the push back against the system was based in a lack of understanding of how it worked, and a lack of trust.
For example, a former prosecutor explaining why he did not ``put much stock'' in risk assessment tools said, ``I'd prefer to look at actual behaviors.
I just didn't know how these tests were administered, in which circumstances, with what kind of data $[\ldots]$ With these tools the output is only as good as the input.
And the input is controversial''~\citep{christin2017algorithms}.
Overall, she concluded that legal professionals ``openly contest the data and methods used to build risk-assessment tools, which they characterize as ``crude'' and ``problematic,'' and criticize the for-profit companies who designed them.
She found that legal professionals questioned why they should follow a completely opaque model over their own professional judgment, noting that ``legal professionals do not see the point.
For better or worse, they trust their own judgment instead.''~\citep{christin2017algorithms}.

These findings suggest that the vision of fairness captured within the COMPAS system does not in any straightforward way tell us something about the fairness of the process as a whole.
The tension between the legislature's desire for some data-driven objectivity to advance fairness and the judges' and bailiffs' sense of the contextual and situational analysis required to advance justice reveals the extent and depth of contests about what fairness requires in practice.\footnote{
Lab experiments find that achieving outcomes perceived as fair requires discussion, not just the establishment of a priori preferences and rules.
See~\citep{lee2017algorithmic}.
} It also highlights the limitations of focusing exclusively on the outputs of the COMPAS system if what we actually care about is the fairness of the pretrial or sentencing processes as a whole.

Whether deploying risk assessments is fair will depend on how we define fairness, the level of abstraction at which we attempt to analyze it, and the particularities of how the assessments are integrated into the criminal justice system.
In the context of the debate around COMPAS, the focus has, for better and worse, been largely on the bias of the instrument itself and on its general opacity (its internals remain a trade secret, even while the broad outlines of its function are straightforward to infer).
But in the context of the criminal justice system, fairness means and demands many things.

\subsection{Applying the Analytic Retrospectively to Unpack the COMPAS Case Study}
We can use our fairness analytic to untangle the strands of the knot in the fairness debate around criminal risk assessment systems and to understand the nature of debate for these systems.
The analytic provides a language and method to elucidate core tensions and dilemmas fueling debates about the fairness implications of these tools.
For concreteness, because it was the situation considered in the acrimonious debate around ProPublica's investigations of COMPAS, we will consider fairness in the use of risk assessments to determine appropriate pre-trial sanctions (i.e., whether arrestees should be detained pre-arraignment or pre-trial or whether they should be releasd on their own recognizance~\cite{pai2019riskassessments}).\footnote{More generally, COMPAS has been considered for use at sentencing as well, and this is the source of the issue in the \emph{Loomis} case~\cite{hlr2017loomis}. Pre-trial sanctions loom large in the discussion of risk assessments, as reducing incarcerated populations is a major goal of criminal justice reform and in many states the bulk of the jail population is pre-sentence. This was the goal, for example, of California's SB10 legislation, which attempted decarceration by eliminating money bail while preserving judges' interests in assessing risk by creating a new pre-trial workflow for arrestees based on risk assessments~\cite{pai2019riskassessments}. Risk assessments are also used in some cases pre-arrest, by officers exercising their discretion to cite, arrest, or refer for charges when encountering citizens in the field, and post-sentencing to determine parole status and sanctions, as well as which prisoners are good candidates for alternative, in-community supervision and support programs.}

First, a \emph{theory of fairness} clarifies both what fairness is expected to do in the world, as well as the moral basis--the underlying beliefs or assumptions--that supports it. At a high level, the various participants in the debates about criminal risk assessments might well agree that respecting each person's individuality and treating them with dignity and respect are relevant purposes of fairness because the risk recidivism tool is in service of the administration of justice in society. That is, while there is high-level agreement about the \emph{purpose} of fairness, the reasons we would want COMPAS to be deployed in fair ways, there is disagreement about the \emph{target} (is fairness in this context about giving similarly situated people similar scores or about how the criminal justice system operates in society and whom it privileges? does it demand robust information and participation rights?) and also the \emph{subject} (Does fairness center on the interest of the accused, or also those of the victim? Do we ascribe fairness to the outputs of the COMPAS model or to the behaviors of the criminal justice system?). Some stakeholders, such as courts and system vendors such as Northpointe in  the debate around COMPAS, see the correct target as rating the relevant risk in the same way for all subjects, privileging a procedural \emph{mechanism} for fairness; other stakeholders, such as ProPublica in the debate around COMPAS and civil rights organizations, view fairness for this tool in light of \emph{consequences}, noting that disparate error rates imply injust treatment for individuals due to circumstances beyond their control (that is, their race); individuals subject to risk assessments instead view fairness at a different \emph{scope}, privileging the justice furnished in their particular case rather than the entire function of the system overall. In the arguments around criminal risk assessments one can see participants offering different justifications for these purposes of fairness, ranging from the inherent sameness of individuals to recognizing past oppression. These differences in justification relate to different positions on the \emph{relationship between theories of fairness and justice}.

This divergence becomes more apparent at the level of \emph{provision} and \emph{protection}.
While all debate participants might agree that a reasonable purpose for fairness is to give each person what they deserve~\cite{velasquez2014justice}, they break down at the level of how to achieve it, that is, in our analytic, at the dimension of target---the concrete thing fairness should deliver, and therefore the appropriate interventions which .
For judges and the criminal justice system more broadly, including the jurisdictions which procure or develop risk assessments, the target can be conceived of as formal equality---blindness to variables such as race and gender---except to the extent they relate to perceived underlying risk levels of subjects as understood through historical data.
For ProPublica, the fairness target should have captured not only the procedural regularity provided by formal equality, but---due to the unequal representation of riskiness of different populations due to historic overpolicing of black and brown communities---also that risk assessments behave similarly in terms of accuracy and error rates across important societal subgroups.
Already, this analysis identifies a tension within demands for fairness, and a tradeoff which must be navigated: the demands of different stakeholders cannot simultaneously be met---the demand of formal equality and procedural regularity cannot be squared mathematically with the demand for equal error rates, precisely because different subjects are situated differently within society~\cite{kleinberg2016inherent, chouldechova2017fair}.
Given the individuality purpose, the target should also include not being detained based on factors beyond an individual's control.
And due to the dignitary purpose, ProPublica's conception of fairness includes protections for information and participation rights (for example access to information about the algorithm as well as the ingested data used to assess risk).
Finally, the judicial context also brings in a particular vision of the relationship between justice and fairness. Specifically, ``Blackstone's Ratio'', the principle of law that it is ``better that ten guilty persons escape than that one innocent suffer'' suggests that justice requires risk assessments to err toward release.

As fairness \emph{relates to society}, many of ProPublica's concerns in the COMPAS case apply to the perspective of a citizen at large: risk assessments should function properly for every member of society (or at least for a wide swath) to preserve overall norms of justice and to protect the social contract~\cite{rawls2009theory,rawls2001justice}. While fairness overlaps substantially with justice in this application context, this observation is nearly vapid without an understanding of the interests served by justice. And yet these are strongly contested within society. The effect on justice of applying risk assessments within the justice system depends on where and how they are applied. For pre-trial risk assessments, fairness can be seen from several \emph{stakeholder viewpoints}, as described in the sections above. The experience of an individual arrestee differs enormously from the perspective of a court clerk, judge, administrator, or journalist. Further, individuals demand not only that decisions are accurate, but that inaccurate decisions can be noted, understood, examined, challenged, and corrected in a timely manner. Finally, most stakeholders are concerned with the principle that assessments be conducted based on relevant and accurate information, supporting the principle of due process and tying the operation of the computer system to the operation of the rest of the justice system.

Finally, it is critical to understand the unit of analysis and the \emph{boundary of ``the system''} when discussing fairness of risk assessment systems. Are we speaking only of the fairness of the scores, of the fairness of the scores as mediated through judicial decision-making, or of overall outcomes with respect to pre-trial detention?\footnote{Other authors refer to this distinction as the ``framing trap'' in understanding fairness~\cite{selbst2019fairness}.} ProPublica's analysis of COMPAS considers primarily the fairness of the scores in isolation~\cite{propublica-bias}. Green and Chen consider the relationship of judges to the scores, as well as issues across the rest of the criminal justice system~\cite{green2019disparate}. Outcomes-oriented measures are the purview of criminal justice policy analysis, and track closely to systems-level concerns. Notably, the risk of unfairness is quite different in a situation where risk assessments are used to select low-risk arrestees for automatic release vs. a robust adversarial hearing as compared to a situation where risk assessments recommend or enforce the automatic detention of high-risk arrestees or are used to deny individuals access to adversarial proceedings. Here, we encounter the issue that fairness of the overall system requires that adversarial hearings which back-stop the correctness of risk assessment outputs must be robust, in the sense that individuals must have their interests vigorously represented, erroneous scores should be demonstrably correctable, and adequate system resources must be available to facilitate this. Also under the heading of where the system's boundary falls comes the question of whether the tools are appropriate to the task: applying a risk assessment designed to predict failure to appear is statistically inappropriate if the question at hand is about the risk of recidivism. Applying a tool designed based on information about prisoners or parolees is inappropriate for making predictions about pre-trial arrestees. The assumptions of the system must reasonably relate to the boundaries of analysis.

In terms of the meaning of \emph{unfairness} in the context of COMPAS, a major point of disagreement is the validity of statistical treatment, which rates individuals based only on a defined set of features as opposed to casuistically~\cite{binns2018s}. Here, jurisdictions have traded off the harms of statistical treatment for efficiency in the administration of justice, but this necessitates access to information and capability to contest erroneous decisions that is not present in COMPAS applications~\cite{hlr2017loomis}. Statistical treatment is especially problematic given observed differences in outcomes that can be attributed to luck egalitarianism~\cite{anderson1999point}

The COMPAS debates serve as an example of the many ways fairness can be defined, analyzed, and operationalized within a sphere of human activity, a sociotechnical system, or a piece of technology.
It reveals the futility of assessing fairness abstractly, if one aims to deliver meaningful solutions in practice.
Going beyond the latent fissure over fairness between the legislature and the judge, fairness in the criminal justice system is both about substantive rules and procedures---thinking about substantive versus procedural fairness also requires thinking about different definitions, visions, and practices of fairness. 
Narrow views of the intrinsic fairness of a tool in isolation (as considered by both ProPublica and Northpointe in their lively debate) may not tell the entire fairness story.

There are other situations in which our analytic provides a useful tool for understanding the contours of debates about the meaning of fairness in a particular application.
For example, the use risk assessment tools for allocating resources of child protective services offices; in Allegheny County, researchers from multiple disciplines are working with policymakers to think about what it means to produce fair outcomes for children and their families~\cite{chouldechova2018case}.
Yet these systems have been criticized as unjust and therefore unfair due to the ways they define risks and harms, and the ensuing ways that cases are screened in or out of the CPS system~\cite{eubanks2018automating}.
Similarly, the New York City Algorithms task force might benefit from tools for communicating about fairness across disciplines as it seeks to leverage the expertise of researchers from multiple disciplines to build consensus and develop both technical and nontechnical interventions to build sociotechnical systems that are more fair \citep{NYCPressOffice2018Mayor}.

\subsection{Applying the Analytic Prospectively}
As shown above the analytic assists in unpacking and understanding disputes about fairness in the COMPAS case.
However, we believe its highest use is to facilitate conversations about the concept of fairness during earlier stages of a project. 
A current example of where the fairness analytic might be productively used is in the design and analysis of pilot studies of risk assessment tools in 16 California counties recently funded by the California legislature as a precursor to the California Money Bail Reform Act (SB 10), which eliminated the state's system of money bail in favor of algorithmic pretrial risk assessments. 
While currently stayed pending the outcome of a 2020 ballot referendum, these pilot projects offer an opportunity to use the fairness analytic to influence how these pilot projects are scoped, designed, and evaluated.
Among other things the pilot projects aim to ``...validate and expand the use of risk assessment tools; and assess any bias.'' \cite{CaliforniaJudicialCouncil2019} 
As we have detailed above, and others have recognized, designing algorithmic tools that align with notions of fairness and justice is complicated \cite{pai2019riskassessments,eaglin2017constructing,cowgill2018impact}.
These tools are part of larger socio-technical systems comprised of legal rules and institutions, organizational processes, and a range of human and technical actors.
Thus analyzing fairness in their design, and assessing it based on their impact, could be facilitated by close attention to the various dimensions identified by the analytic. 

The structure of SB10 which allocates responsibility for setting policies and managing the development of validated risk assessment tools to the Judicial Council but grants local courts substantial discretion over the creation and implementation of the risk assessment instruments themselves (\S\S1320.24.; 1320.26(a)), raises questions about how fairness will be theorized, protected, provided, and measured. 
For example, as others have noted, overarching policies set by the Judicial Council could produce particular kinds of unfairness given differing local conditions \cite{solow2019institutional}.
As in the COMPAS debate discussed above, if base rates vary across localities applying globally defined thresholds, as currently required (\S1320.25.), it will produce particular kinds of unfairness. 
The analytic could assist in rethinking how general policies and local realities combine to produce particular kinds of fairness. 
It could support conversations that translate between theories and provisioning in specific systems in specific local contexts. 
SB 10 maintains judges as the final arbiter of pretrial release or detention decisions (See, e.g., \S1320.20(f)) yet whether judges will continue to exercise the same level of discretion over such decisions when pretrial risk assessment reports are available is uncertain. 
To the extent that maintaining the judge's centrality is essential to fairness, the analytic could promote reflection on what mechanisms beyond the law might support such judge's independence---training, interface design, policies, etc. 
Relatedly, it could prompt conversations about when the fairness of pilot programs is measured---for example, should the existence of the policy of judge primacy, or empirical analysis of levels of judge independence after the tool is introduced, be the measure of fairness? 
One co-author hopes to use the fairness analytic to influence the design and evaluation of these 16 pilot programs and in future work hopes to report on the experience.  
\section{Discussion and Reflections}

Our definitional overview of conceptions of fairness highlights how terms are used in multiple ways both within and between different disciplinary and professional communities, and provides the groundwork for using these terms in ways that comprise shared understanding. Building on that, the Fairness Analytic offers a bridging tool to bring distinct disciplines into shared conversations about the purpose, justifications, and mechanisms for advancing fairness despite domain specific vocabularies with often latent assumptions and politics, and the ongoing contestations about what fairness demands in varied contexts.

We offer the Fairness Analytic as a heuristic or tool for future interdisciplinary research collaborations and discussions. 
Rather than trying to converge on a single definitive definition of fairness, our approach highlights the diversity of fairness conceptions that may be profitably brought into practice. 
The analytic we offer may be viewed as a boundary negotiation object---an object used to ``record, organize, explore and share ideas; introduce concepts and techniques; create alliances; create a venue for the exchange of information; augment brokering activities; and create shared understanding''~\citep{bietz2009collaboration}.
Specifically, the analytic is a \emph{compilation artifact} that reflects the divergent way terms are defined in computer science, social sciences, and law.
Compilation artifacts ``bring two or more communities of practice into alignment just long enough to develop a shared and mutually agreeable understanding of a problem and to pass crucial information from one community of practice to another''~\citep{lee2005between}.
As such, it provides a useful tool for exploring the problem space and can help us collectively consider the definitions of concepts we have to choose from, and negotiate shared understandings that help us bring some order to our conversations, and select the concepts best aligned with contextually determined goals.

Using the lens of infrastructures while curating the multiple dimensions of this term, fairness, attuned us to the power and politics involved in using, defining, and deploying terms. In a situated context, a particular discipline's definition or conceptualization of fairness may carry more power or weight and be seen as the obvious or ``natural'' definition. 
Yet not recognizing other conceptualizations, and their context of analysis, carries potential risks in missing out on other important factors. For instance, focusing on the fairness of COMPAS in terms of fair outcomes might obscure the need or desire to also have fair processes. Both outcomes and processes are potential targets of fairness, and often conceptions of fairness require both. The way fairness is conceptualized when discussing outcomes will likely be different than when discussing processes. 
The definitional overview of terms helps surface this multiplicity of definitions and concepts to the forefront when working with the values of fairness, while the analytic provides a tool for discussion of multiple concepts of fairness. 
It is our hope that future researchers and practitioners reference the definitional overview and use the analytic as a translation tool to speak and collaborate across disciplinary boundaries, and also as a reflexive tool to be cognizant of these political and power differences when moving among different uses of this value. 
The analytic may also be used as a way to support meaningful contestation, as a way to make clear the differences among assumptions and meanings that stakeholders brings to discussions of fairness.

The lens of infrastructures also suggest paying attention to ongoing maintenance practices of sociotechnical systems as moments for values to (re)emerge as salient for debate and design. 
In this spirit, the definitional overview and are analytic are not a ``complete'' artifacts; the definitional overview does not provide an exhaustive specification of every possible definition for the set of values across every possible discipline, nor is the analytic a complete specification of every dimension across which fairness might be dissected. 
Rather, the coverage of these terms and dimensions provides a specific starting point to support and expand researchers' ongoing capacity to converse and collaborate across boundaries. 
Together the definitional overview and analytic also help clarify different contexts of usage, so that these ongoing moments of debate can lead to ``meaningful contestation'' \citep{mulligan2016privacy} rather than talking past one another. 

We also note that the definitional overview and analytic should be considered as parts of a broader ecosystem of social and technical infrastructures among this interdisciplinary community interested in fairness in (socio)technical systems. 
The usefulness of these tools will emerge in relation to other sociotechnical infratructuring practices, such as: the creation and maintenance of spaces for discussion and collaborations including workshops, conferences, research groups, and discussion forums; training and teaching within multiple communities of practice; or funding mechanisms for projects in these domains. 
The future contributions of the definitional overview and analytic may be in the work it helps us organize, or the heightened form of organization in the field that working with and through them might yield. 
Future work may also investigate the multiplicity of related terms accountability, transparency, and explainability.
As Hunsinger et al. write, ``infrastructure is indeed a fundamentally relational concept; it emerges for people in practice, connected to activities and structures''~\cite{hunsinger2010international}.
The definitional overview and analytic are an effort at infrastructuring for the community of scholars and practitioners studying fairness, accountability, transparency, explainability, and related concepts as they apply to computer systems; as such it is part object, part process, and---we hope---part practice.

\section{Conclusion}
The concept of fairness is vast and ambiguous, and differently used across disciplines.
Research and practice aimed at advancing these properties in sociotechnical systems powered by machine learning algorithms requires cross-disciplinary engagement.
 A shared definitional overview and Fairness Analytic that helps map variations on fairness within and across disciplines and domains is a necessary, but insufficient, element of the infrastructure to support such work.
By enabling researchers from different backgrounds to recognize when terms are being used differently, and potentially settle on shared understandings for specific projects, our conceptual analytic can help researchers collaborate and sustain political debates about what fairness is for and what it demands of sociotechnical systems.
A more careful understanding of fairness will lead to better cross-disciplinary collaborations both studying and building systems.
 Finally, infrastructuring is a political task.
The taxonomies we construct and the dimensions we reveal in the analytic, and how we use and deploy them, will shape the research agenda and outputs created.
It will also determine how well they meet the needs of society.
\section*{Acknowledgements}
The authors give thanks to the participants at the 2018 ACM FAT* conference and the 2018 Neural Information Processing Symposium, and the 2018 Privacy Law Scholars Conference for feedback on earlier iterations of this work.

The authors also thank the members of the UC Berkeley Algorithmic Fairness and Opacity working group, members of the DARPA Information Science and Technology advisory board, and participants in the Simons Institute for the Theory of Computing Summer 2019 Cluster on Fairness for insightful discussions of fairness.

This work has benefited specifically from discussions with and feedback provided by (in alphabetical order)  Alicia Solow-Niederman, Annette Zimmerman, Andrew Selbst, Brenda Leong, Emmanuel Moss, Jen Gennai, Miranda Bogen, and Suresh Venkatsubramanian.

Research for this article has been supported by generous funding from the US NSF INSPIRE SES1537324, NSF DGE 1752814, the Berkeley Center for Law and Technology, and a generous gift from Microsoft.
\newpage
\appendix
\section{An Analytic for Applying Contested Conceptions of Fairness in Computer Systems}

\begin{table}[h!]
  \noindent\begin{tabularx}{\textwidth}{||X|X|X|X||}
    \hline \hline
  \textbf{\mbox{Dimension of} \mbox{Fairness}} & \textbf{\mbox{Description of} \mbox{Dimension}} & \textbf{\mbox{Example of} \mbox{Dimension}} & \textbf{\mbox{Interrogation} \mbox{Questions}} \\ \hline \hline
  \multicolumn{4}{||c||}{\textsc{Dimensions of Theory}}\\ \hline \hline
  Purpose & That which fairness provides or contributes to those protected. &
  \vspace{-6pt}
  \begin{itemize}[leftmargin=*, noitemsep, topsep=0pt, label={$\circ$}]
  \item Individuality (i.e., not viewed as a statistic)
  \item Dignity (ability to participate)
  \item Respect (presentation of self/recognition of self)
  \item Empathy (recognition of unequal starting points, capacities, etc.)
  \end{itemize}& What is fairness for?\\ \hline
  Justification & That which justifies fairness, i.e., \emph{fairness is justified because of X} &
  \vspace{-6pt}
  \begin{itemize}[leftmargin=*, noitemsep, topsep=0pt, label={$\circ$}]
  \item Historical Oppression (Racism, Sexism, etc.)
  \item Inherent sameness
  \item Diminishing impact of luck/uncontrolled benefits \& risks
  \item Social Welfare
  \item Self development
  \item Incentives for X
  \item Rewards for X
  \item Reparations
  \end{itemize} & Why should this be fair?\\ \hline
  Contrast Concept & That which contrasts to fairness, i.e., \emph{that which is fair is mutually exclusive with that which is X} &
  \vspace{-6pt}
  \begin{itemize}[leftmargin=*, noitemsep, topsep=0pt, label={$\circ$}]
  \item Inequality
  \item Bias
  \item Discrimination
  \item Racism, Sexism
  \end{itemize} & What's not fair?\\ \hline
  Exemplar & The prototype of fairness, i.e., \emph{fairness is best illustrated by X} & ``one person, one vote'' & What's an example?\\ \hline \hline
\end{tabularx}
\caption{Dimensions of theory for contests over fairness.}
\end{table}

\begin{table}[p!]
  \noindent\begin{tabularx}{\textwidth}{||X|X|X|X||}
    \hline \hline
  \textbf{\mbox{Dimension of} \mbox{Fairness}} & \textbf{\mbox{Description of} \mbox{Dimension}} & \textbf{\mbox{Example of} \mbox{Dimension}} & \textbf{\mbox{Interrogation} \mbox{Questions}} \\ \hline \hline
  \multicolumn{4}{||c||}{\textsc{Relationship to Society}}\\ \hline \hline
  Relationship to Justice & The extent to which fairness overlaps with or supports justice. &
  \vspace{-6pt}
  \begin{itemize}[leftmargin=*, noitemsep, topsep=0pt, label={$\circ$}]
  \item Distributive justice
  \item Retributive justice
  \item Compensatory justice
  \end{itemize} & How does justice relate to fairness?\\ \hline
  Stakeholder Viewpoint & Fair as seen from whose point of view? (This person might not be the subject of fairness analysis.) & Stakeholder viewpoint &  Fair as seen from whose point of view? (This person might not be the subject of fairness analysis.)\\ \hline
  Boundary of the System & The scope of the system being evaluated for fairness. &
  \vspace{-6pt}
  \begin{itemize}[leftmargin=*, noitemsep, topsep=0pt, label={$\circ$}]
  \item Technical System
  \item Technical system and user
  \item Technical system and user and organization
  \item Technical system and user and organization and norms/regulations
  \end{itemize} & What is the (socio-)technical system that is being made fair?\\ \hline \hline
  \multicolumn{4}{||c||}{\textsc{Dimensions of Unfairness}}\\ \hline \hline
  Action & That which contributes or constitutes unfairness &
  \vspace{-6pt}
  \begin{itemize}[leftmargin=*, noitemsep, topsep=0pt, label={$\circ$}]
  \item Unequal treatment/discrimination
  \item Unequal outcomes
  \item Unequal process
  \item Stereotyping
  \item Statistical Treatment
  \end{itemize} & What violates fairness?\\ \hline
  Offender & Actor(s) violating fairness, i.e., \emph{fairness violated by agent X}. &
  \vspace{-6pt}
  \begin{itemize}[leftmargin=*, noitemsep, topsep=0pt, label={$\circ$}]
  \item Government
  \item Business entity
  \item Individual
  \end{itemize} & Who violates fairness?\\ \hline
  Consequences of unfairness & What harms result from a defecit of fairness? &
  \vspace{-6pt}
  \begin{itemize}[leftmargin=*, noitemsep, topsep=0pt, label={$\circ$}]
  \item Loss of benefits
  \item Loss of dignity
  \item Loss of autonomy
  \end{itemize} & For stakeholders of the system, what are the consequences of failing to realize fairness?\\ \hline
\end{tabularx}
\caption{Dimensions of unfairness for contests over fairness.}
\end{table}

\begin{table}[pt!]
  \noindent\begin{tabularx}{\textwidth}{||X|X|X|X||}
    \hline \hline
  \textbf{Dimension of \mbox{Fairness}} & \textbf{Description of \mbox{Dimension}} & \textbf{Example of \mbox{Dimension}} & \textbf{Interrogation \mbox{Questions}} \\ \hline \hline
  \multicolumn{4}{||c||}{\textsc{Dimensions of Protection}}\\ \hline \hline
  Target & The ideal end state toward which fairness aspires. At a high level, this could be substantive or procedural. &
  \vspace{-6pt}
  \begin{itemize}[leftmargin=*, noitemsep, topsep=0pt, label={$\circ$}]
  \item Formal equality (blind to all other variables)---to each person an equal share;
  \item Need-based allocation---to each person according to individual need;
  \item Effort-based allocation---to each person according to individaul effort;
  \item Social contribution---to each person according to societal contribution;
  \item Merit-based allocation---to each person according to merit;
  \item Information and participation rights
  \item Accurate and robust representation
  \end{itemize} & What should fairness provide?\\ \hline
  \multirow{2}{0.22\textwidth}{Subject (and, in relation to who/what?)} & Actor(s) or Entity(ies) to whom fairness is provided. &
  \vspace{-6pt}
  \begin{itemize}[leftmargin=*, noitemsep, topsep=0pt, label={$\circ$}]
  \item Individual
  \item Social Groups
  \item Roles
  \end{itemize} &
  Fairness is at stake for whom or what?\\ 
  & Fairness is often used comparatively, requiring the construction of categories along some attribute or set of attributes.
  & 
  & What properties or attributes are being made fair? What groups are being compared? Granularity? \\\hline
\end{tabularx}
\caption{Dimensions of protection for contests over fairness.}
\end{table}
\hfill \break
\hfill \break

\begin{table}[t!]
  \noindent\begin{tabularx}{\textwidth}{||X|X|X|X||}
    \hline \hline
  \textbf{\mbox{Dimension of} \mbox{Fairness}} & \textbf{\mbox{Description of} \mbox{Dimension}} & \textbf{\mbox{Example of} \mbox{Dimension}} & \textbf{\mbox{Interrogation} \mbox{Questions}} \\ \hline \hline
  \multicolumn{4}{||c||}{\textsc{Dimensions of Provision}}\\ \hline \hline
  Provider & Actor(s) charged with being fair or avoiding unfairness. &
  \vspace{-6pt}
  \begin{itemize}[leftmargin=*, noitemsep, topsep=0pt, label={$\circ$}]
  \item Government
  \item Business entity
  \item Technology
  \item Individuals
  \end{itemize} & Who or what is supposed to behave fairly or avoid unfair behavior?\\ \hline
  Mechanism & Modalities used to support fairness. &
  \vspace{-6pt}
  \begin{itemize}[leftmargin=*, noitemsep, topsep=0pt, label={$\circ$}]
  \item Legal regulations
  \item Technical design
  \item Business processes
  \item Education
  \item Norms
  \end{itemize} & How is fairness operationalized?\\ \hline
  Implementer & Actor(s) tasked with operationalizing fairness through chosen modalities. &
  \vspace{-6pt}
  \begin{itemize}[leftmargin=*, noitemsep, topsep=0pt, label={$\circ$}]
  \item Lawyers
  \item Engineers
  \item Product Managers
  \item Designers
  \item Professional associations
  \item Educators
  \end{itemize} & Who brings fairness into practice?\\ \hline \hline
  \multicolumn{4}{||c||}{\textsc{Dimensions of Context}}\\ \hline \hline
  Social Practice & That wherein fairness applies, i.e., \emph{a situation, a field, a site, a model}. &
  \vspace{-6pt}
  \begin{itemize}[leftmargin=*, noitemsep, topsep=0pt, label={$\circ$}]
  \item Hospital or University
  \item Nation-State or Globally
  \end{itemize} & What is the context of fairness?\\ \hline
  Scope & Extent of application of fairness, i.e., \emph{fair should be applied at scope X}. &
  \vspace{-6pt}
  \begin{itemize}[leftmargin=*, noitemsep, topsep=0pt, label={$\circ$}]
  \item Universally as strict rule
  \item Casuistically as per-case
  \end{itemize} & How widely does fairness apply?\\ \hline
  Time & The time period(s) over which fairness is measured and observed, i.e., \emph{fairness applies for span X of time}. &
  \vspace{-6pt}
  \begin{itemize}[leftmargin=*, noitemsep, topsep=0pt, label={$\circ$}]
  \item Before decisions are made
  \item When decisions are made
  \item After decisions are made
    \end{itemize} & When is fairness observed? When do we measure what is fair? Is this a static process or a dynamic one?\\ \hline
\end{tabularx}
\caption{Dimensions of provision and context for contests over fairness.}
\end{table}

%
\bibliographystyle{ACM-Reference-Format}
\bibliography{lexicon-cscw}

%
\end{document}